\pdfoutput=1
\documentclass{my-paper}

\title{Mapping Partisan Fault Lines Within \glstext*{dao}s\thanks{This
is a preprint of a published article~\citep{lloyd-et-al-25}.}}

\author{Thomas~Lloyd\thanks{\email{tomlloyd1992@gmail.com}}}

\author{Daire~\'O~Broin\thanks{\email{daire.obroin@setu.ie},
\orcid{0000-0002-2886-5546}}}

\author{Martin~Harrigan\thanks{\email{martin.harrigan@setu.ie},
\orcid{0000-0002-6069-7001}}}

\affil{\gls*{setu}, \acrlong*{roi}}

\date{}

\begin{document}

\maketitle

\begin{abstract}
  \glsstar{}{dao}{s} can fragment when partisan communities emerge
within their governance structures, leading to organisational splits
known as \textquote[][.]{forks}  We present a method to detect these
emerging communities by analysing on-chain voting behaviour before
fragmentation occurs.  Our approach extracts voting events from
governance smart contracts, constructs voter matrices encoding
participation patterns, and applies pairwise dissimilarity analysis to
quantify voting divergence between addresses.  We visualise these
relationships using multidimensional scaling and identify partisan
communities through $k$-means clustering with silhouette score
optimisation.  Using Nouns \gls*{dao} as a case study, a protocol that
has experienced multiple documented forks, we demonstrate that
addresses that would later fork cluster together months before actual
fragmentation events.  Our analysis of \num{330} proposals spanning
from contract deployment to the first major fork shows that
\qty{90}{\percent} of fork addresses cluster together in the final
\num{44} proposals, compared with only \qty{47}{\percent} in
randomised data.  These results indicate that partisan communities
can be detected and visualised through on-chain governance analysis,
providing retrospective evidence of emerging divisions before they
cause organisational fragmentation.

\end{abstract}

\section{Introduction}\label{sec:introduction}

\glsstar{}{dao}{s} have experienced rapid growth in both number and
participation since 2019~\citep{bellavitis-et-al-23}, with governance
decisions now affecting billions of \dUSD[post-between=]{} in digital
assets.  Unlike traditional organisations with known stakeholders,
\glsstar{}{dao}{s} are governed by pseudonymous token-holding
addresses, creating challenges for understanding community dynamics
and predicting organisational stability.  While
\glsstar{}{dao}{s} often present themselves as unified communities,
the inherent nature of governance, involving disagreement, debate, and
competing interests, can lead to the formation of distinct
sub-communities that may ultimately fragment the organisation.

The fragmentation of \glsstar{}{dao}{s} through
\textquote[][,]{forking} where dissenting members create new instances
of the protocol, is an important aspect of decentralised governance.
Forking divides communities and resources and undermines the
legitimacy and effectiveness of the original organisation.  Early
detection of partisan communities that might lead to fragmentation is
essential for maintaining long-term protocol stability and effective
governance.

This paper presents a method for identifying and visualising
partisan communities within \glsstar{}{dao}{s} by analysing on-chain
voting behaviour.  Our approach transforms blockchain governance data
into spatial representations that reveal behavioural alignment
patterns among participants.  By examining the voting history of
addresses across multiple proposals, we can detect the emergence of
cohesive sub-communities before they manifest as organisational
splits.

Our method consists of several stages: we extract voting events from
on-chain governance smart contracts and construct voter matrices
encoding participation and support patterns across all proposals.  We
then apply pairwise dissimilarity analysis to quantify voting
divergence between addresses based on their shared voting history.
These dissimilarity scores are projected into two dimensions using
\gls*{mds}, creating spatial representations where proximity
indicates voting alignment.  Finally, we apply $k$-means clustering
with silhouette score optimisation to identify distinct partisan
communities, and we validate the results against a randomised
baseline that isolates genuine alignment from coincidental agreement.

We demonstrate our approach on Nouns \gls*{dao}\footnote{Nouns
\gls*{dao} homepage: \url{https://nouns.wtf/}}, an
\glstext*{nft}-based governance system on Ethereum whose documented
forks provide a behavioural ground truth against which our clusters
can be validated.  Our analysis spans from the deployment of the
governance contract through the period preceding the first major
fork.  Our results show that addresses that would later fork can be
detected within distinct clusters months before the actual
fragmentation event occurs.  As proposals approach a fork, voter data
shows near-unanimous clustering of future fork participants, while
randomised data exhibits no such pattern.

The remainder of this paper is organised as follows:
\Cref{sec:related-work} reviews related work on \gls*{dao} governance
behaviour, multidimensional scaling for political analysis, and
decentralisation measurement frameworks.  \Cref{sec:prelim-analysis}
applies an initial version of the method to \gls*{uk} parliamentary
voting data, where party affiliation supplies a similar ground truth
in a non-pseudonymous setting and serves as a sanity check on the
dissimilarity and \gls*{mds} pipeline before we apply it to
pseudonymous blockchain data.  \Cref{sec:nouns-dao} describes Nouns
\gls*{dao} and explains why we chose it as the case study.
\Cref{sec:method} details our multi-stage method for transforming
blockchain data into community visualisations.  In \cref{sec:results}
we present our validation procedures and our analysis of Nouns
\gls*{dao}, including \gls*{mds} visualisations and clustering
results compared against randomised baselines.  We conclude in
\cref{sec:conclusions} with implications for \gls*{dao} governance.

\section{Related Work}\label{sec:related-work}

We organise related work into three areas: \gls*{dao} communities and
governance behaviour, multidimensional scaling for political analysis,
and decentralisation measurement frameworks.

\subsection{\gls*{dao} Communities and Governance Behaviour}

The pseudonymous nature of \glsstar{}{dao}{s} creates challenges for
community analysis, as participants must reach consensus on
multi-billion \dUSD[post-between=]{} decisions without traditional
trust mechanisms.  Research in this area has moved from basic
controversy metrics to coalition detection methods that account for
pseudonymity and delegation.

\citet{sun-et-al-22-1} identified five distinct coalitions within
MakerDAO~\citep{sky-xx} through clustering analysis, showing how
voting blocs can be detected and their impact on protocol decisions
measured.
Similarly, \citet{dupont-23} applied graph-based deep learning to
detect Sybil clusters, revealing that \qtyrange{2}{5}{\percent} of
voter addresses exhibit suspicious behavioural patterns.

\citet{austgen-et-al-23} introduced the theoretical framework of
\emph{voter-block entropy} (VBE), which treats aligned addresses as
unified voting entities rather than individual accounts.  While their
approach remains theoretical, it provides the conceptual foundation
for our practical method of identifying partisan communities through
historical voting behaviour.  Their work on Dark \glsstar{}{dao}{s},
which attempt to buy votes opaquely, highlights the importance of
transparency in governance analysis.

Recent studies have incorporated off-chain community sentiment
analysis.  \citet{quan-et-al-24} analysed Discord discussions across
six \glsstar{}{dao}{s} using natural language processing, finding
generally positive sentiment in governance channels.  However, this
approach cannot capture the verified, immutable voting behaviour that
on-chain analysis provides.

\subsection{Multidimensional Scaling for Political Analysis}

The application of \gls*{mds} to political voting analysis has proved
effective in traditional parliamentary systems, providing the
methodological foundation for our \gls*{dao} community detection
approach.

\citet{mickevicius-et-al-14} demonstrated the use of MDS for
parliamentary voting analysis in Lithuania.  They showed that
clustering algorithms can reliably identify opposing political
parties.  Their work established \gls*{mds} as a suitable method for
visualising ideological alignment in voting systems.

\citet{amelio-et-al-15} analysed Italian parliamentary votes using
temporal segmentation, dividing their dataset into six-month intervals
to capture evolving political dynamics.  Their approach of tracking
faction formation over time directly informs our method of analysing
\gls*{dao} voting patterns across sequential proposals.  They showed
that \gls*{mds} visualisations can reveal the emergence of new
political coalitions before they become formally recognised.

\subsection{Decentralisation Measurement and \gls*{dao} Assessment}

Understanding \gls*{dao} decentralisation has become important as
regulators grapple with how to classify these novel organisational
structures.  Research has revealed concerning centralisation patterns
across multiple \glsstar{}{dao}{s}.

\citet{fritsch-et-al-22} found that fewer than \num{19} addresses
control majority voting power in three major \glsstar{}{dao}{s},
highlighting the concentration of governance influence.  This finding
shows why our address-level analysis remains relevant despite apparent
power concentration: coalition formation among minority holders can
still influence majority decisions.

\citet{feichtinger-et-al-23} performed a related analysis across
\num{23} Ethereum \glsstar{}{dao}{s}, revealing widespread
centralisation in voting power distribution alongside low
participation rates and significant costs from governance
transactions.  Their work on \textquote{pointless governance
transactions} shows the importance of filtering for genuine,
sustained participation in our analysis.

The same authors' systematisation of \gls*{dao} attack
vectors~\citep{feichtinger-et-al-24} identified behavioural factors as
the primary source of governance exploits across \num{28} real-world
incidents.  Their findings show the importance of understanding
community dynamics to predict and prevent governance failures.

\citet{axelsen-et-al-22} developed the TIGER framework for measuring
decentralisation across five dimensions, applying design-science
methodology to create verifiable decentralisation metrics.  Their work
addresses regulatory concerns about what constitutes
\textquote{sufficient decentralisation} but focuses on structural
rather than behavioural measures.

Historical perspectives on governance crises provide additional
context.  \citet{reijers-et-al-18} analysed the resolution to The DAO
hack in 2016 as a \textquote[][,]{state of exception} revealing how
private interests can override established governance rules during
emergencies.  This analysis highlights the fragility of decentralised
governance systems and the importance of detecting partisan divisions
before they reach crisis points.

Our work synthesises insights from all three areas: we apply
\gls*{mds} techniques from political analysis to \gls*{dao}
governance behaviour,
providing a practical tool for measuring the kind of community
fragmentation that decentralisation frameworks seek to prevent.

\section{Preliminary Political Analysis}\label{sec:prelim-analysis}

Detecting community alignment from voting records is a well-studied
problem in legislative analysis.  \citet{mickevicius-et-al-14}
analysed voting behaviour in the Lithuanian Parliament and found that
multidimensional scaling visualised opposing parties almost
perfectly, with cluster assignments matching political parties at
\qty{68}{\percent} accuracy.  The appeal of legislative data for this
kind of analysis is that the governance record is part of the public
parliamentary record, and party members are expected to vote in
alignment with the party's policy objectives, providing a known
ground truth against which community-detection methods can be
evaluated.

A clear example is the \gls*{uk} parliamentary system, where
political parties formalise cohesive voting through the party whip:
each \gls*{mp} is instructed on how to vote in each parliamentary
proposal.  The \gls*{uk} government also releases all governance data
to the public.  Because the major parties operate the whip system and
the data is freely available, we used a subset of the voting records
from the House of Commons to apply an initial version of our method
to detect similarities between voters and visualise the result in an
\gls*{mds} graph.

Unlike the Nouns \gls*{dao} study presented later in this paper,
which evaluates the complete community identification method, this
preliminary analysis applies only the pairwise dissimilarity measure
and the \gls*{mds} visualisation.  It serves as a sanity check,
showing that meaningful visualisations of governance communities can
be generated directly from voting history, with the known party
affiliations of \glsstar{}{mp}{s} providing a ground truth by which
the resulting \gls*{mds} visualisation can be assessed.

House of Commons governance data is released per proposal and
includes each \glsstar{}{mp}{'s} name, party affiliation,
constituency, vote, and proxy member.  Our data collection began on
8th May 2024 and continued to 9th October 2024.  We unified the
proposals into a single voter matrix and determined voter alignment
scores through pairwise dissimilarity comparisons of all voters; the
method is described in detail in \cref{sec:method}.  Once the
dissimilarity matrix had been computed, we projected the data into
two dimensions using \gls*{mds}\@.  During the period of analysis the
\gls*{uk} held a general election on 4th July 2024; consequently, the
set of \glsstar{}{mp}{s} eligible to participate in governance
changed.  We therefore generated \gls*{mds} graphs both for the
entirety of our data and for the post-election period alone.
\Cref{fig:mp-mds} shows the post-election projection of \gls*{mp}
voting alignment.  \Cref{fig:mp-mds-all} projects dissimilarity
across both epochs, before and after the election.

\begin{figure}
  \centering
  \myincludegraphics[width=\linewidth]{png/figures/mp-mds}
  \caption{Parliamentary \gls*{mds} graph for
    Proposals~\numrange{1827}{1844}.  Each vertex is an \gls*{mp},
    coloured by political-party affiliation.  The distance between
    vertices reflects dissimilarity in voting behaviour: greater
    distance indicates greater dissimilarity.}\label{fig:mp-mds}
\end{figure}

\begin{figure}
  \centering
  \myincludegraphics[width=\linewidth]{png/figures/mp-mds-all}
  \caption{Parliamentary \gls*{mds} graph across all proposals.  The
    distinct grouping of same-party vertices reveals two separate
    epochs, before and after the election.}\label{fig:mp-mds-all}
\end{figure}

In \cref{fig:mp-mds}, each vertex is coloured by its political-party
affiliation, with proximity indicating greater voting alignment and
separation indicating misalignment.  The Conservatives, Labour, and
the Liberal Democrats occupy distinct regions of the graph.  This is
consistent with our expectation that the whip system aligns voting
behaviour within each party, and validates the dissimilarity and
\gls*{mds} pipeline on data with a known ground truth.  The seemingly
random placement of the \textquote{Other} category reflects its
composition: a mix of smaller parties and independents whose voting
behaviour is not tied to a single party line and may align with
either the left-leaning or right-leaning blocs.

\Cref{fig:mp-mds-all} is harder to interpret.  The Conservative and
Labour vertices each split into two distinct groups.  This is an
artefact of the way we determine dissimilarity for voters who have
never participated in the same governance proposal: when there is no
overlap between two voters' activity, we treat them as maximally
dissimilar.  Parliamentary data segments naturally into epochs
between general elections, with limited cross-overs through
by-elections, and \glsstar{}{mp}{s} who sat only before the election
therefore never share a proposal with those who sat only after.
\gls*{dao} governance has no equivalent boundary: voters may enter or
exit the system at any time and without a formal process, so the same
dissimilarity convention will produce noise rather than coherent
groupings unless the active voters are filtered, which we do in
\cref{sec:active-voters}.

The preliminary analysis confirms that our approach can visualise
partisan communities in a setting with a known ground truth.
Translating the method to \gls*{dao} governance introduces challenges
that are absent in parliamentary data.  The chief difficulty is the
inability to verify community detection directly: blockchain
addresses are pseudonymous, and no equivalent of party affiliation is
available against which clusters can be validated.  We therefore
restrict our subsequent analysis to a single \gls*{dao} that provides
a behavioural ground truth.  We selected Nouns \gls*{dao} as our test
case because it has previously undergone forks, in which a subset of
the membership separated from the main \gls*{dao} to create a new
instance.  The forks identify, after the fact, which addresses left,
providing a proxy for partisan-community membership.

\section{Nouns \gls*{dao}}\label{sec:nouns-dao}

Nouns \gls*{dao} describes itself as an experimental attempt to
improve the formation of on-chain avatar communities.  Noun avatars
are represented as on-chain \glstext*{nft}s, each with a distinct
combination of visual traits.  The \gls*{dao} generates revenue by
auctioning a new Noun \glstext*{nft} every \num{24} hours on
Ethereum: bidders pay in ether, and after \num{24} hours the highest
bidder receives the \glstext*{nft} and the proceeds are transferred
to the Nouns \gls*{dao} treasury.

The treasury is operated by the \gls*{dao} itself.  Nouns implements
a fork of the Compound governance framework, with governance taking
place on-chain and voting rights granted to Nouns \glstext*{nft}
holders at one vote per Noun.  The \gls*{dao} regularly holds
proposals on allocations from the treasury, and has funded a wide
range of projects, from staking positions in Lido Finance to a global
pizza party costing roughly \dUSD{42000}.  Disagreement surrounding
this kind of spending appears to have contributed to a group within
the \gls*{dao} initiating the first Nouns \gls*{dao}
fork~\citep{kubinec-23}.

The first fork saw \num{472} Nouns exit the original \gls*{dao} to
form a new instance, with the treasury split proportionally and the
corresponding share of ether transferred to the new \gls*{dao}\@.
The forked \gls*{dao} did not rebuild a working community; instead, a
new \emph{rage quit} mechanism within it allowed Noun owners to
liquidate their \glstext*{nft}s for a proportional share of the new
treasury.  In effect, this gave those who forked a route to exit the
Nouns ecosystem entirely.  Nouns \gls*{dao} has since undergone two
further forks.

We selected Nouns \gls*{dao} as the case study for mapping partisan
communities for two reasons.  First, forking a \gls*{dao} is an
extreme form of conflict resolution, and its presence in Nouns
\gls*{dao} is itself indicative of partisan divisions within the
membership.  Second, the forks provide a ground truth: they identify,
after the fact, which addresses ultimately separated from the
\gls*{dao}, allowing us to examine the relationship between an
address's community alignment and its future likelihood of joining a
fork.

\Cref{fig:dao-governance-nouns-dao} illustrates the Nouns \gls*{dao}
governance scenario.  Two communities of actors each hold the Nouns
governance token and vote in Nouns \gls*{dao}\@.  Community~1
exercises yes~votes and Community~2 exercises no~votes.  Detecting
these opposing communities from on-chain voting data \emph{before}
the fork occurs is the goal of the method presented in this paper.

\begin{figure}
  \centering
  \myincludegraphics[width=0.7\linewidth]{tikz/dao-governance/dao-governance-nouns-dao}
  \caption{Internal community governance in Nouns \gls*{dao}\@.  Two
    communities hold the Nouns token and vote in opposing directions
    in Nouns \gls*{dao}\@.  Community~1 exercises yes~votes and
    Community~2 exercises
    no~votes.}\label{fig:dao-governance-nouns-dao}
\end{figure}

\section{Method}\label{sec:method}

To identify partisan communities within \glsstar{}{dao}{s}, we propose
a multi-stage method that bridges on-chain data analysis with spatial
and statistical validation.  The following subsections cover each
stage (\crefrange{sec:blockchain-data}{sec:cluster-analysis}).

Briefly, the method begins with \emph{blockchain data acquisition},
where we gather raw transaction logs and event data from
\glstext*{evm}-based \glsstar{}{dao}{s} through \glstext*{rpc} calls.
The data is structured into a \emph{voter matrix}, encoding voting
behaviour while accounting for abstentions and non-participation.  We
perform a \emph{community friction assessment} to quantify overall
voter discord across proposals.  This acts as a sieve, prioritising
\glsstar{}{dao}{s} with significant partisan dynamics over those with
consistently unanimous voting.  Then we identify \emph{active voters}
by selecting addresses with a sufficient level of participation
within a sliding proposal window; this limits subsequent analysis to
engaged participants, reducing noise from sporadic voters.  The core
of our method lies in \emph{pairwise dissimilarity computation},
where we measure voting divergence between addresses by comparing
their shared voting history.  This produces a dissimilarity matrix
per proposal, capturing how often two voters opposed each other
relative to their co-participation.  We visualise these matrices
using \emph{multidimensional scaling} (\glstext*{mds}), mapping
addresses into 2D space where proximity reflects voting alignment.
Finally, we identify partisan communities using \emph{dynamic
clustering}, selecting cluster counts through \emph{silhouette
scoring} and testing robustness by comparing genuine voting data
against randomised voting patterns.

Our method is governance-framework agnostic and can be applied
broadly across \glsstar{}{dao}{s}.  We explain it using Nouns
\gls*{dao} (\cref{sec:nouns-dao}) as a running example: voting power
in Nouns \gls*{dao} is derived from \glstext*{nft} ownership, and its
documented history of forks provides a ground truth against which we
can compare our results.

\subsection{Blockchain Data Acquisition}\label{sec:blockchain-data}

We collected on-chain transaction and event logs from \glstext*{evm}
blockchains (Ethereum and Arbitrum) through \glstext*{rpc} calls,
using a self-hosted Ethereum node and Infura.  We identified
governance smart contracts for six \glsstar{}{dao}{s}: Lido,
Compound, Tornado Cash, Nouns, Uniswap, and Arbitrum.  In previous
work, we implemented a systematic approach to identifying governance
smart contracts~\citep{lloyd-et-al-24}.  However, for this work, our
focus is a deeper analysis of a smaller number of \glsstar{}{dao}{s}
and it sufficed to identify the contracts by inspecting the projects'
documentation and transaction history.  For each \gls*{dao}, we
determined the relevant event signature(s) by analysing historical
transactions and filtering on the voting events emitted by the
governance contract.  The events contain proposal identifiers, voter
addresses, and the direction of each vote (support, opposition, or
abstention).  We collected event logs across the entire operational
history of each \gls*{dao}, from the block in which the governance
contract was deployed, up to and including block \num{22575000} (May
2025).  The Nouns \gls*{dao} governance contract\footnote{Contract
Address: \texttt{0x6f3E6272A167e8AcCb32072d08E0957F9c79223d}} was
deployed in block \num{12985453} (August 2021).

\subsection{Voter Matrix Construction}\label{sec:voter-matrix}

\begin{figure}
  \centering
  \myincludegraphics[width=\linewidth]{png/figures/nouns-dao-voter-matrix}
  \caption{A sample from the voter matrix for Nouns \gls*{dao} with
    ten voting addresses (the rows) and nine proposals (the columns).
    The cell values represent either \textquote{Yes} votes (\num{1}s
    in green), \textquote{No} votes (\num{0}s in red), or all other
    cases (\num{-1}s in grey).}\label{fig:nouns-dao-voter-matrix}
\end{figure}

After parsing, each vote log entry contained the voting address, the
proposal ID, and a support value.  We transformed this data into a
voter matrix, illustrated in \cref{fig:nouns-dao-voter-matrix}, where
rows correspond to voter addresses, columns represent proposal IDs,
and entries denote support: \num{1} for \textquote{Yes}, \num{0} for
\textquote[][,]{No} and \num{-1} to represent all other cases.  The
raw data distinguishes five states: \num{1} (Yes), \num{0} (No),
\num{-1} (no voting power and no vote), \num{2} (had voting power and
abstained), and \num{99} (had voting power but did not vote).  We
subsumed the latter three states (\numlist{-1; 2; 99}) into a single
\num{-1} value to simplify our analysis; each represents an absence
of a yes or no vote, and the binary distinction is what our
subsequent analysis requires.  Determining voting eligibility
precisely is non-trivial: an address might have held tokens at the
time of a proposal but only become eligible retroactively due to
delayed vesting schedules, delegation mechanics, or other temporal
constraints, and accurate eligibility tagging requires checking
historical token balances at specific block heights.  To construct
the per-proposal voter lists used in later stages, we extracted
addresses with non-negative support values (\num{0} or \num{1}) for
each proposal.  Although \glsstar{}{dao}{s} often use a support value
of \num{2} to denote abstentions that contribute to quorum, we
adopted a binary classification (for/against) so that the voter
matrix aligns with standard voting-analysis frameworks.

More formally, if $A = \{a_1, \ldots, a_n\}$ is the set of
addresses that participated in a \gls*{dao} and $P = \{p_1, \ldots,
p_m\}$ is the set of proposals voted on by the \gls*{dao},
then the voter matrix $V$, with $n$ rows and $m$ columns, is defined
such that $v_{ij} \in \{0, 1, -1\}$ indicates the action taken by
address $a_i$ on proposal $p_j$:
\begin{equation*}
  V =
  \begin{bmatrix}
    v_{11} & v_{12} & \cdots & v_{1m} \\
    v_{21} & v_{22} & \cdots & v_{2m} \\
    \vdots & \vdots & \ddots & \vdots \\
    v_{n1} & v_{n2} & \cdots & v_{nm}
  \end{bmatrix}
\end{equation*}

For Nouns \gls*{dao}, we constructed a matrix with \num{629} unique
voter addresses (rows) and \num{330} proposals (columns) spanning a
period from the deployment of the governance smart contract to the
first Nouns \gls*{dao} fork in block \num{18144239} (September 2023).
The proposal IDs were numbered \numrange{1}{362} inclusive; we
excluded \num{32} cancelled proposals that had no voting activity,
leaving $362 - 32 = 330$ proposals.

\subsection{Community Friction Assessment}\label{sec:community-friction}

\begin{figure}
  \centering
  \myincludegraphics[width=\linewidth]{png/figures/dao-disagreement-stacked}
  \caption{A percent stacked bar chart showing disagreement across all
    proposals for six \glsstar{}{dao}{s} at the voter address level.
    Lido \gls*{dao} voters are unanimous (\qty{0}{\percent}
    disagreement) or exhibit low disagreement (in the $(0, 20)\%$
    range) for almost all of their proposals whereas Nouns \gls*{dao}
    voters exhibit medium disagreement (in the $[20, 40)\%$ range)
    and high disagreement (in the $[40, 50]\%$ range) for over
    \qty{30}{\percent} of their
    proposals.}\label{fig:dao-disagreement-stacked}
\end{figure}

To prioritise \glsstar{}{dao}{s} with meaningful partisan dynamics, we
quantified community-level discord using two metrics derived directly
from the voter matrix (see \cref{sec:voter-matrix}): static
disagreement percentages per proposal and rolling average disagreement
percentages over sequential proposals.

For static disagreement analysis, we categorise proposals by the
proportion of voters opposing the winning outcome.  The thresholds are
unanimous (\qty{0}{\percent} disagreement), low ($(0, 20)\%$), medium
($[20, 40)\%$), and high ($[40, 50]\%$).  These align with standard
benchmarks for social consensus in collective decision-making
studies.  As shown in \cref{fig:dao-disagreement-stacked}, Nouns
\gls*{dao} exhibits medium-to-high disagreement in
\qty{30}{\percent} of proposals, a stark contrast to Lido
\glsstar{}{dao}{'s} near-unanimous voting patterns.  This
quantitative disparity motivates our selection of Nouns \gls*{dao}
as a high-friction case study, particularly given its documented
fork in September 2023.

\begin{figure}
  \centering
  \myincludegraphics[width=\linewidth]{png/figures/dao-disagreement-rolling}
  \caption{A plot of the rolling average of disagreement for six
    \glsstar{}{dao}{s} using a \num{10}-proposal sliding window.  The
    $x$-axis contains the proposal IDs, which serve as a proxy for
    time.  The $y$-axis shows the level of disagreement.  Lido
    \gls*{dao} remains close to unanimity throughout, while Nouns
    \gls*{dao} exhibits sustained conflict leading up to its first
    fork.}\label{fig:dao-disagreement-rolling}
\end{figure}

The rolling average metric (see \cref{fig:dao-disagreement-rolling})
reveals temporal trends in partisan behaviour.  By calculating the
mean disagreement percentage across a \num{10}-proposal window (moving
forward one proposal at a time), we detect persistent voting
divisions in Nouns \gls*{dao}, where disagreement spikes and sustained
volatility preceded its fork.  This metric serves two purposes: (1)
identifying \glsstar{}{dao}{s} with persistent rather than transient
conflict, and (2) pinpointing temporal clusters of contentious
governance that warrant further analysis (see
\cref{sec:cluster-analysis}).

These friction metrics inform our subsequent dynamic clustering
analysis.  \glsstar{}{dao}{s} with medium/high disagreement exceeding
\qty{20}{\percent} (see \cref{fig:dao-disagreement-stacked}) and
rolling averages exceeding \qty{15}{\percent} (see
\cref{fig:dao-disagreement-rolling}) are flagged for further analysis,
while low-friction \glsstar{}{dao}{s} such as Lido are excluded.  This
filtering ensures our clustering experiments (see
\cref{sec:cluster-analysis}) focus on cases where conflict is evident.

\subsection{Active Voter Identification}\label{sec:active-voters}

To address the sparsity of the voter matrix caused by fluctuating
participation in \glsstar{}{dao}{s}, we compute a voter participation
list.  This distinguishes \gls*{dao} voting from parliamentary
voting, where persistent cohorts dominate.  In \glsstar{}{dao}{s},
voting eligibility depends on dynamic factors such as token
ownership, delegation, and time-based vesting, leading to variable
active-voter sets across proposals.  As a result, many addresses have
minimal or no voting activity, introducing noise into community
detection.

Our method calculates participation frequency using a sliding proposal
window.  For each proposal, we define a window containing the proposal
and a fixed number of preceding proposals.  For Nouns \gls*{dao}, we
used a window of size \num{10}.  We chose this size because it
captures a specific moment in the \glsstar{}{dao}{'s} history: it
provides enough context for community formation to be observed, while
still responding to changes once communities have formed.  A larger
window would delay the observation of community evolution, both into
alignment and out of it.  For Proposal~\num{20}, the window included
Proposals~\numrange{11}{20} inclusive.  For each voter address in
each window, we compute their participation percentage:
\begin{equation*}
  \textrm{Participation \unit{\percent}} = \frac{\textrm{\# of
  proposals in which the address voted}}{\textrm{\# of proposals
  in the window}}
\end{equation*}

Addresses with a participation percentage below a fixed threshold
(\qty{40}{\percent} for Nouns \gls*{dao}) were dropped from the
filtered voter matrix.  We chose \qty{40}{\percent} because it
ensures that voters participated in enough proposals for the later
pairwise dissimilarity comparison: it provides sufficient overlap for
meaningful comparisons and reduces the noise from pairs of voters who
never voted together, while still allowing cases of no shared voting
history to be represented in the data.  This ensures subsequent
analysis focuses on voters with sustained engagement.  Early
proposals (Proposals~\numrange{1}{9} for Nouns \gls*{dao}) used
shorter windows (for example, the window for Proposal~\num{2}
included only Proposals~\numlist{1;2}), increasing voter counts until
sufficient historical data was accumulated.

More formally, let $w \in \mathbb{N}$ denote the proposal window size
and $\tau \in [0, 1]$ the participation threshold.  For each proposal
$p_j \in P$, we define its sliding window $W_j
\subseteq P$ as:
\begin{equation*}
  W_j = \left\{ p_{\max(1, j - w + 1)}, p_{\max(1, j - w +
  2)}, \dots, p_j \right\}.
\end{equation*}

For each address $a_i \in A$, we define its participation
percentage $\pi_{ij}$ in $W_j$ as:
\begin{equation*}
  \pi_{ij} = \frac{1}{|W_j|} \sum_{p_k \in W_j}
  \mathbf{1}_{\{v_{ik} \geq 0\}},
\end{equation*}
where $\mathbf{1}$ is the indicator function.  We use $|W_j|$ rather
than $w$ in the denominator so that the shorter windows used at the
start of the dataset (for example, $|W_1| = 1$) are normalised
correctly.  The filtered voter matrix $V'$ retains only rows $a_i$
with $\pi_{ij} \geq \tau$, and the columns corresponding to the
proposals in $W_j$.

The filtered voter matrix concentrates the analysis on the addresses
with sustained engagement, for which partisan alignment is meaningful
to measure, and serves as the input to the pairwise dissimilarity
computation (see \cref{sec:pairwise-dissimilarity}).

\subsection{Pairwise Dissimilarity
Computation}\label{sec:pairwise-dissimilarity}

The core of our method is the pairwise dissimilarity computation,
which measures the voting divergence between voter addresses.  It
quantifies how often two addresses vote in opposition relative to
their shared participation across proposals, producing a
dissimilarity matrix for each proposal that captures the voting
alignment within the \gls*{dao}\@.  We chose vote-based pairwise
dissimilarity as the metric because it normalises disagreement by the
number of proposals on which each pair of addresses cast valid votes.

For each proposal $p_j \in P$, we compute pairwise
dissimilarity scores between all active voter addresses within the
proposal's sliding window $W_j$ (as defined in
\cref{sec:active-voters}).  Let $A_j \subseteq A$
denote the set of active addresses for proposal $p_j$, filtered by the
participation threshold $\tau$.  For any two addresses $a_i, a_k \in
A_j$, their dissimilarity score $d^j_{ik}$ is calculated as:
\begin{equation*}
  d^j_{ik} =
  \begin{cases}
    \dfrac{\sum_{p_{\ell} \in W_j}
      \mathbf{1}_{\{v_{i\ell}, v_{k\ell} \geq 0 \,\land\, v_{i\ell} \neq
      v_{k\ell}\}}}{\sum_{p_{\ell} \in W_j}
      \mathbf{1}_{\{v_{i\ell}, v_{k\ell} \geq 0\}}}
      & \text{if } \exists\, p_{\ell} \in W_j :
        v_{i\ell}, v_{k\ell} \geq 0 \\
    1 & \text{otherwise.}
  \end{cases}
\end{equation*}

The numerator counts the number of proposals within $W_j$
where both addresses cast valid votes (that is, $v_{i\ell}, v_{k\ell}
\geq 0$) but voted in opposition to each other.  The denominator
represents the total number of proposals where both addresses
participated with valid votes.  When two addresses have no overlapping
voting history within the window, we assign the maximum dissimilarity
score of \num{1}, as pairs with no shared votes cannot show voter
alignment; this occurred in less than \qty{0.6}{\percent} of all pairs
across all proposals.  This formulation ensures that dissimilarity
scores are normalised to the range $[0, 1]$, where \num{0} indicates
complete alignment (addresses always vote together when both
participate) and \num{1} indicates complete opposition (addresses
never agree when both participate).

\begin{algorithm}
  \caption{\textsc{ComputeDissimilarity} generates pairwise
  dissimilarity matrices for all proposals.}\label{alg:compute-dissim}
  \begin{algorithmic}[1]
    \Function{ComputeDissimilarity}{$V$, $A$, $w$, $\tau$}
      \State $R \gets \{\}$ \Comment{Dictionary to store results}
      \ForAll{$p_j \in P$}
        \State $W_j \gets \{p_{\max(1, j-w+1)}, \ldots, p_j\}$
          \Comment{Sliding window}
        \State $A_j \gets \{a_i \in A : \pi_{ij}
          \geq \tau\}$ \Comment{Active addresses}
        \State $V' \gets V[A_j, W_j]$
          \Comment{Filtered voter submatrix}
        \State $n \gets |A_j|$
        \State $D^j \gets \mathbf{0}_{n \times n}$
          \Comment{Initialise dissimilarity matrix}
        \For{$i \gets 1$ \textbf{to} $n-1$}
          \For{$k \gets i+1$ \textbf{to} $n$}
            \State $\text{valid} \gets \{p_{\ell} \in W_j :
              v_{i\ell}, v_{k\ell} \geq 0\}$
            \State $\text{shared} \gets |\text{valid}|$
            \If{$\text{shared} > 0$}
              \State $\text{opposing} \gets |\{p_{\ell} \in \text{valid} :
                v_{i\ell} \neq v_{k\ell}\}|$
              \State $d^j_{ik} \gets \text{opposing} / \text{shared}$
            \Else
              \State $d^j_{ik} \gets 1$ \Comment{No shared voting history}
            \EndIf
            \State $d^j_{ki} \gets d^j_{ik}$ \Comment{Symmetric assignment}
          \EndFor
        \EndFor
        \State $R[p_j] \gets \{\text{addresses}:
          A_j, \text{matrix}: D^j\}$
      \EndFor
      \State \Return $R$
    \EndFunction
  \end{algorithmic}
\end{algorithm}

\Cref{alg:compute-dissim} outlines the computational procedure.  For
each proposal $p_j$, we construct the sliding window $W_j$
and identify the set of active addresses $A_j$ that meet the
participation threshold.  We then extract the relevant submatrix
$V'$ from the voter matrix $V$ and compute pairwise
dissimilarity scores.

\begin{figure}
  \centering
  \myincludegraphics[width=\linewidth]{png/figures/nouns-dao-dissimilarity-matrix}
  \caption{A sample from the dissimilarity matrix for Nouns \gls*{dao}
    Proposal~\num{300}.  The rows are the addresses and the columns
    are the same addresses (abbreviated).  A score of \num{1}
    represents complete dissimilarity; a score of \num{0} represents
    complete similarity.}\label{fig:nouns-dao-dissimilarity-matrix}
\end{figure}

The resulting dissimilarity matrix $D^j$ for proposal $p_j$
is an $n \times n$ symmetric matrix where $n = |A_j|$.  Each
entry $d^j_{ik}$ represents the dissimilarity between addresses $a_i$
and $a_k$, with the diagonal entries being zero (complete
self-similarity).  \Cref{fig:nouns-dao-dissimilarity-matrix} shows a
sample from the dissimilarity matrix for Nouns \gls*{dao}
Proposal~\num{300}.

For Nouns \gls*{dao}, this process generated \num{329} dissimilarity
matrices: one per proposal except the first, which has no preceding
proposals from which to compute a window.  Each matrix captures the
voting alignment within the \gls*{dao} at that point in time.  These
matrices serve as the foundation for subsequent multidimensional
scaling visualisation (see \cref{sec:mds}) and dynamic clustering
analysis (see \cref{sec:cluster-analysis}).

\subsection{Spatial Visualisation via \glstext*{mds}}\label{sec:mds}

To visualise the alignment structure captured by our dissimilarity
matrices, we use \gls*{mds} to project the high-dimensional voting
relationships into 2D space.  \gls*{mds} finds an embedding in which
the Euclidean distances approximate the pairwise dissimilarities, so
addresses with similar voting behaviour appear close together.  As
demonstrated in \cref{sec:prelim-analysis}, \gls*{mds} can visualise
political alignment from voting records alone; we apply it here to
the pairwise dissimilarity matrices to identify and visualise
alignment between voting addresses.

For each proposal $p_j$ with dissimilarity matrix $D^j$, we
apply classical \gls*{mds} to obtain a 2D embedding $\mathbf{X}^j \in
\mathbb{R}^{n \times 2}$ where each row $\mathbf{x}^j_i$ represents
the coordinates of address $a_i$ in the visualisation space.  The
\gls*{mds} algorithm minimises the stress function:
\begin{equation*}
  \text{Stress} = \sqrt{\frac{\sum_{i<k} (d^j_{ik} -
    \lVert\mathbf{x}^j_i - \mathbf{x}^j_k\rVert_2)^2}{\sum_{i<k}
  (d^j_{ik})^2}}
\end{equation*}
where $\lVert\cdot\rVert_2$ denotes the Euclidean distance in the 2D
embedding space.  This ensures that addresses with low dissimilarity
scores appear close together in the visualisation, while highly
dissimilar addresses are positioned far apart.

To maintain temporal continuity across proposals, we use the embedding
coordinates from proposal $p_{j-1}$ as initialisation for the
\gls*{mds} algorithm at proposal $p_j$.  This prevents arbitrary
rotations and reflections between consecutive visualisations,
enabling coherent tracking of voter movement over time.  We performed
\gls*{mds} using the \texttt{scikit-learn} library with the following
parameters: metric \gls*{mds} with Euclidean distance, maximum of
\num{300} iterations, and convergence tolerance of \num{e-6}.  The
embedding positions $\mathbf{X}^j$ were stored for subsequent
clustering analysis (see \cref{sec:cluster-analysis}).

\begin{figure}
  \centering
  \myincludegraphics[width=\linewidth]{png/figures/nouns-dao-mds-with-gt}
  \caption{\glstext*{mds} visualisation for Nouns \gls*{dao}
    Proposal~\num{334} with ground-truth colouring.  Each vertex
    represents a voter address, with positions determined by pairwise
    dissimilarity scores.  Red vertices indicate addresses that
    participated in a subsequent fork, while blue vertices represent
    addresses that remained in the original
    \gls*{dao}\@.}\label{fig:nouns-dao-mds-with-gt}
\end{figure}

Nouns \gls*{dao} provides a natural ground truth for validating our
visualisation approach.  The \gls*{dao} has undergone three forks
since deployment, each initiated when a subset of \glstext*{nft}
holders (Noun owners) signalled their intent to establish a new
\gls*{dao} instance via on-chain transactions.  We identified fork
participants by analysing \glstext*{nft} transfers to the new fork
smart contracts, creating a binary classification of addresses as
either \textquote{forkers} or \textquote[][.]{stayers}

\Cref{fig:nouns-dao-mds-with-gt} shows the \gls*{mds} visualisation
for Proposal~\num{334}, where red vertices represent addresses that
participated in a subsequent fork and blue vertices represent those
that remained.  The spatial clustering of the vertices shows that our
dissimilarity-based approach captures behavioural divisions that
manifest in real-world governance actions.  This ground truth shows
that \gls*{mds} translates abstract voting patterns into
interpretable spatial relationships that can be used for automated
partisan community detection.

\subsection{Dynamic Cluster Analysis}\label{sec:cluster-analysis}

\begin{figure}
  \centering
  \myincludegraphics[width=\linewidth]{png/figures/nouns-dao-mds-with-k-means}
  \caption{\glstext*{mds} visualisation for Nouns \gls*{dao}
    Proposal~\num{334} with $k$-means clustering.  Note the overlap
    with the ground-truth clusters in
    \cref{fig:nouns-dao-mds-with-gt}.}\label{fig:nouns-dao-mds-with-k-means}
\end{figure}

Once each proposal has an \gls*{mds} embedding, we cluster the
embeddings to identify partisan groups.  This stage transforms the
spatial patterns visible in our visualisations into quantifiable
community structures that can be analysed and validated.  We chose
$k$-means clustering because it groups addresses by proximity in the
\gls*{mds} embedding, identifying communities with similar voting
behaviour and allowing us to quantify changes within communities over
time.

For each proposal $p_j$, we perform $k$-means clustering on the 2D
coordinates $\mathbf{X}^j$ obtained using \gls*{mds} (see
\cref{sec:mds}).  Since the number of active voters varies across
proposals and partisan structures evolve over time, we dynamically
determine the optimal number of clusters rather than use a fixed
value.  We use the \emph{silhouette score} to select the optimal
cluster count $k^*$ for each proposal.  The silhouette score $s(i)$
for voter $i$ is calculated as:
\begin{equation*}
  s(i) = \frac{b(i) - a(i)}{\max\{a(i),\, b(i)\}}
\end{equation*}
where $a(i)$ is the average distance from voter $i$ to all other
voters in the same cluster, and $b(i)$ is the average distance from
voter $i$ to all voters in the nearest neighbouring cluster.  The
overall silhouette score for a clustering configuration is the mean of
individual silhouette scores across all voters.  Silhouette scoring
accommodates the dynamic nature of community structures: it
determines the most appropriate number of communities within each
proposal window, letting us observe when voters consolidated into a
few cohesive groups or fragmented across many.

We evaluate cluster counts from $k = 2$ to $k = 5$ inclusive and
select the configuration with the highest silhouette score, as shown
in \cref{fig:nouns-dao-silhouette-coefficient}.  We restrict the
search to this range because partisan structures in
\glsstar{}{dao}{s} typically present as a small number of aligned
groups rather than as highly fragmented communities.  In
supplementary tests up to $k = 10$, \qty{22}{\percent} of clusterings
exceeded five groups overall, but fewer than \qty{3}{\percent} did so
after Proposal~\num{319}.  Hence, restricting to $k \leq 5$ has a
negligible impact on detecting partisan structures.

\begin{figure}
  \centering
  \myincludegraphics[width=\linewidth]{png/figures/nouns-dao-silhouette-coefficient}
  \caption{Silhouette coefficient scores for cluster counts $k = 2$ to
    $k = 5$ on Nouns \glsstar{}{dao}{'s}
    Proposal~\num{334}.}\label{fig:nouns-dao-silhouette-coefficient}
\end{figure}

As a preliminary validation, we can use the ground truth provided by
Nouns \glsstar{}{dao}{'s} documented forks.
\Cref{fig:nouns-dao-mds-with-k-means} shows the clustering results for
Proposal~\num{334}, where $k^* = 2$.  The clusters closely match the
ground truth division shown in \cref{fig:nouns-dao-mds-with-gt}, with
forking addresses predominantly clustered together and distinct from
addresses that remained in the original \gls*{dao}\@.

\subsection{Limitations}\label{sec:limitations}

The method includes design choices at every stage that may influence
the resulting community structures.  Our choices of a sliding window
of \num{10} proposals, a participation threshold of
\qty{40}{\percent}, vote-based pairwise dissimilarity, \gls*{mds},
$k$-means clustering, and silhouette scoring over cluster counts up
to \num{5} were based on their appropriateness in the context of this
study; alternative parameter values and clustering approaches were
not evaluated.  In future work, we will perform a sensitivity
analysis to evaluate the robustness of the detected communities and
implement alternative methods for measuring dissimilarity and
clustering the resulting data.

\section{Results}\label{sec:results}

Our analysis of Nouns \gls*{dao} governance data spans from the
deployment of its governance contract at block \num{12985453} (8th
August 2021) through Proposal~\num{362} at block \num{18050498} (2nd
September 2023).  This time frame captures the complete pre-fork
governance history, providing an insight into how partisan communities
emerged and solidified before the \glsstar{}{dao}{'s} fragmentation.

\subsection{Visualisation of Community Structure}

The \gls*{mds} visualisation in \cref{fig:nouns-dao-mds-with-gt}
shows a spatial representation for Proposal~\num{334}, which we
selected for detailed analysis due to its high concentration of
participating fork addresses.  The graph maps 2D positions of
governing addresses based on their voting alignment, where proximity
indicates similar voting behaviour and distance reflects increasing
voting dissimilarity.

A clear bimodal structure emerges, with two distinct clusters
separated along the horizontal axis.  The rightmost cluster contains
addresses that subsequently participated in one of Nouns
\glsstar{}{dao}{'s} three forks (in red), while the leftmost cluster
contains addresses that remained in the original \gls*{dao} (in
blue).  This spatial segregation shows that our dissimilarity-based
approach captures underlying differences in observed voting behaviour
that align with the eventual fragmentation of the \gls*{dao}\@.  The
left-right clustering pattern persists across all subsequent
proposals, suggesting that once partisan identities formed, they
remained through the pre-fork period.

\subsection{Quantitative Clustering Analysis}

Applying $k$-means clustering with silhouette score optimisation to
the \gls*{mds} coordinates confirms our visual observations.  For
Proposal~\num{334}, the algorithm selected $k^* = \num{2}$ as the
optimal number of clusters, in agreement with the binary fork/no-fork
ground truth.  Fourteen of the fifteen fork addresses were assigned
to one cluster, with only a single fork address placed in the other.
The clustering algorithm, operating solely on geometric proximity
without knowledge of future fork participation, reconstructed the
partisan divide that would later lead to the \glsstar{}{dao}{'s}
fragmentation.

\subsection{Temporal Evolution of Partisan Behaviour}

Our analysis shows a shift in fork addresses' governance engagement
over time.  During the early period (Proposals~\numrange{1}{256}),
fork addresses participated sporadically, with an average of only
\num{1.86} addresses voting per proposal.  Beginning with
Proposal~\num{257} (22nd March 2023), their participation increased.
In the final \num{91} non-cancelled proposals (Proposals~\numrange{257}{362},
of which \num{15} were cancelled), at least four fork addresses
participated in every vote, with average participation jumping to
\num{8.27} addresses per proposal.

This suggests that addresses that would later fork the \gls*{dao}
increased their governance involvement as observable divisions
deepened.  While non-forking addresses also increased their
participation (from \num{28.86} to \num{39.93} average participants),
the proportional increase was much smaller, indicating that
engagement among the addresses that would later fork increased
disproportionately as the fork approached.

\subsection{Consolidation into Partisan Communities}

\begin{figure}
  \centering
  \myincludegraphics[width=\linewidth]{png/figures/nouns-dao-forked-clusters-stacked}
  \caption{A stacked area chart showing the assignment of forked
    addresses in the Nouns \gls*{dao} to $k$-means clusters across
    Proposals~\numrange{257}{362}.  The addresses gradually coalesce
    into a single cluster as the fork
    approached.}\label{fig:nouns-dao-forked-clusters-stacked}
\end{figure}

\Cref{fig:nouns-dao-forked-clusters-stacked} shows the temporal
consolidation of fork addresses into a unified partisan community.
The stacked area chart tracks cluster assignments across
Proposals~\numrange{257}{362}, revealing how initially fragmented fork
addresses gradually coalesced into a single, dominant cluster as the
\gls*{dao} approached its breaking point.  This process shows that
partisan communities do not emerge suddenly but form through
sustained governance participation.  Fork addresses become
increasingly unified in their cluster assignment, suggesting that
their observed voting behaviour became progressively more aligned
over time through repeated participation in governance.

\subsection{Validation Against Randomised Data}

\begin{table}
  \centering
  \caption{Forked cluster dominance and average clustering
    comparison.  For each proposal range, the table reports the
    average number of $k$-means clusters and the percentage of forked
    addresses assigned to the dominant cluster, alongside the
    minimum, maximum, and mean values obtained from a shuffled
    baseline of \num{100} randomised
    iterations.}\label{tab:forked-cluster-summary}
  \begin{tabular}{llrrrr}
    \toprule
    \textbf{Range} & \textbf{Metric} & \textbf{Value} &
    \makecell{\textbf{Shuff.}\\\textbf{Min.}} &
    \makecell{\textbf{Shuff.}\\\textbf{Max.}} &
    \makecell{\textbf{Shuff.}\\\textbf{Avg.}} \\
    \midrule
    \multirow{2}{*}{\makecell{Proposals\\\numrange{1}{362}}}
    & Avg \# Clusters & \num{2.77} & \num{3.14} & \num{3.57} & \num{3.36} \\
    & \unit{\percent} in Forked Cluster & \num{72.20} & \num{57.70} &
    \num{67.64} & \num{62.64} \\
    \midrule
    \multirow{2}{*}{\makecell{Proposals\\\numrange{257}{362}}}
    & Avg \# Clusters & \num{2.48} & \num{3.23} & \num{4.05} & \num{3.66} \\
    & \unit{\percent} in Forked Cluster & \num{79.36} & \num{44.60} &
    \num{59.31} & \num{50.40} \\
    \midrule
    \multirow{2}{*}{\makecell{Proposals\\\numrange{319}{362}}}
    & Avg \# Clusters & \num{2.14} & \num{3.06} & \num{4.14} & \num{3.62} \\
    & \unit{\percent} in Forked Cluster & \num{90.96} & \num{41.66} &
    \num{59.69} & \num{47.36} \\
    \midrule
    \multirow{2}{*}{\makecell{Proposals\\\numrange{349}{362}}}
    & Avg \# Clusters & \num{2.08} & \num{2.50} & \num{4.67} & \num{3.87} \\
    & \unit{\percent} in Forked Cluster & \num{90.28} & \num{34.82} &
    \num{63.24} & \num{46.99} \\
    \bottomrule
  \end{tabular}
\end{table}

\Cref{tab:forked-cluster-summary} compares genuine voting data with
\num{100} randomised iterations in which participants and outcomes
were preserved but vote assignments shuffled (seeds \numrange{0}{99}),
isolating the effect of genuine alignment from random behaviour.

The results are as follows.  First, genuine voter data produces
fewer clusters per proposal (\num{2.77} average) than
randomised data (\num{3.36} average), indicating that genuine voting
behaviour exhibits more coherent community structure than would occur
by chance.  This gap widens as we focus on periods closer to the fork,
with the final \num{14} proposals showing just \num{2.08} clusters on
average versus \num{3.87} in randomised data.

Second, the percentage of fork addresses concentrated within a
single cluster differs sharply between genuine and randomised data.
Across all proposals, \qty{72.20}{\percent} of fork addresses cluster
together in genuine data, compared with only \qty{62.64}{\percent} in
randomised data.  This differential becomes more pronounced in
the pre-fork period:  in the final \num{44} proposals,
\qty{90.96}{\percent} of fork addresses cluster together in genuine
data, against just \qty{47.36}{\percent} in randomised data.

Our results show that the partisan communities that ultimately
fragmented Nouns \gls*{dao} could be detected, in retrospect, before
the fork occurred.  Addresses that later participated in forks
exhibited progressively more aligned voting behaviour months before
the decision to fork was enacted.  These findings suggest that
governance data may contain early indicators of emerging partisan
divisions.  However, this evidence is retrospective; a prospective
prediction or monitoring system would require further validation
across additional \glsstar{}{dao}{s} and false-positive cases.

\section{Conclusions}\label{sec:conclusions}

\glsstar{}{dao}{s} enable open, pseudonymous participation in
significant governance decisions, yet this openness can produce
partisan divisions that are often difficult to identify but can
culminate in disruptive outcomes such as forks.

We proposed a method to map such divisions by analysing on-chain
governance data.  It starts with a voter matrix representing a
\glsstar{}{dao}{'s} governance activity across its proposals and voter
addresses.  We measure the overall friction within the community based
on the outcomes of governance proposals.  We explore this friction at
the address level, filtering the data to include active voting
participants within the \gls*{dao}\@.  We segment proposals into a
rolling window of ten proposals to capture how voting behaviour
evolves throughout the \glsstar{}{dao}{'s} lifecycle, and consider
active voters to be those who participated in \qty{40}{\percent} of
proposals within the window; this filter removes addresses with
minimal engagement and reduces noise in the data.  We then conduct a
pairwise dissimilarity analysis on the active voting addresses within
each proposal window, comparing the binary votes of each pair of
addresses, and visualise the alignment using \gls*{mds}\@.  Once
projected, we identify clusters using $k$-means clustering and select
the optimal configuration using silhouette scoring.  This framework
transforms individual voting records into a coherent visual
representation of voter alignment for each proposal in the
\glsstar{}{dao}{'s} history.

We use Nouns \gls*{dao} as a case study, as its documented forks
provide a ground truth for validating our method.  We show that
addresses that later participated in a fork are increasingly grouped
into a single cluster, with over \qty{90}{\percent} aligned just
before the split.  Moreover, the average number of clusters falls to
roughly two by this stage, indicating two partisan communities.  We
validate these results by randomising voter assignments while
preserving participation and outcomes; across \num{100} iterations of
randomised voter assignments, on average \qty{47}{\percent} of forked
addresses clustered together and the average number of clusters rose
above four.  The alignment between the detected communities and the
forking outcome provides behavioural validation of the method against
a documented fork within a single \gls*{dao}, rather than universal
validation across all \glsstar{}{dao}{s}\@.  The fork provides an
external label for fragmentation, but it does not reveal private
motives, ideology, or deliberate coordination.

Beyond confirming that internal communities exist within
\glsstar{}{dao}{s}, we have shown how our method can automate
community detection from raw blockchain log events.  The degree to
which communities disagree can be quantified and visualised, exposing
partisan divisions within \glsstar{}{dao}{s}\@.  We found that
addresses belonging to a particular cluster had an increased
likelihood of participating in subsequent \gls*{dao} forks.  This is
significant because analysis at the community level can provide
insight into the subsequent behaviour of individual addresses.  The
insight is particularly important in the context of forking, which
can destabilise and delegitimise a \glsstar{}{dao}{'s} governance
system.  If left undetected and unresolved, partisan divisions can
fragment a \gls*{dao} and negatively impact its stability.  Our
method provides a means to monitor these dynamics and a foundation
for early-warning systems, although prospective use would require
validation beyond the retrospective evidence presented here.  In
future work, we will extend our analysis to other \glsstar{}{dao}{s}
that have undergone a fork and perform the sensitivity analysis
described in \cref{sec:limitations}.

\bibliographystyle{my-abbrvnat}
\bibliography{partisan-lines}

\begin{thebibliography}{16}
\providecommand{\natexlab}[1]{#1}
\providecommand{\url}[1]{\texttt{#1}}
\expandafter\ifx\csname urlstyle\endcsname\relax
  \providecommand{\doi}[1]{doi: #1}\else
  \providecommand{\doi}{doi: \begingroup \urlstyle{rm}\Url}\fi

\bibitem[Amelio and Pizzuti(n.d.)]{amelio-et-al-15}
A.~Amelio and C.~Pizzuti.
\newblock Mining and Analyzing the Italian Parliament: Party Structure and
  Evolution.
\newblock In \emph{Recommendation and Search in Social Networks}, pages
  249--279. Springer, n.d.
\newblock ISBN 9783319364803.
\newblock \doi{10.1007/978-3-319-14379-8_12}.

\bibitem[Austgen et~al.(2023)Austgen, F{\'a}brega, Allen, Babel, Kelkar, and
  Juels]{austgen-et-al-23}
J.~Austgen, A.~F{\'a}brega, S.~Allen, K.~Babel, M.~Kelkar, and A.~Juels.
\newblock DAO Decentralization: Voting-Bloc Entropy, Bribery, and Dark DAOs.
\newblock Technical report, 2023.

\bibitem[Axelsen et~al.(2022)Axelsen, Jensen, and Ross]{axelsen-et-al-22}
H.~Axelsen, J.~R. Jensen, and O.~Ross.
\newblock When is a DAO Decentralized?
\newblock \emph{Complex Systems Informatics and Modeling Quarterly},
  176\penalty0 (31):\penalty0 51--75, 2022.
\newblock \doi{10.7250/csimq.2022-31.04}.

\bibitem[Bellavitis et~al.(2023)Bellavitis, Fisch, and
  Momtaz]{bellavitis-et-al-23}
C.~Bellavitis, C.~Fisch, and P.~P. Momtaz.
\newblock The Rise of Decentralized Autonomous Organizations (DAOs): A First
  Empirical Glimpse.
\newblock \emph{Venture Capital: An International Journal of Entrepreneurial
  Finance}, 25\penalty0 (2):\penalty0 187--203, 2023.

\bibitem[DuPont(2023)]{dupont-23}
Q.~DuPont.
\newblock New Online Communities: Graph Deep Learning on Anonymous Voting
  Networks to Identify Sybils in Polycentric Governance.
\newblock Technical report, 2023.

\bibitem[Feichtinger et~al.(2023)Feichtinger, Fritsch, Vonlanthen, and
  Wattenhofer]{feichtinger-et-al-23}
R.~Feichtinger, R.~Fritsch, Y.~Vonlanthen, and R.~Wattenhofer.
\newblock The Hidden Shortcomings of (D)AOs --- An Empirical Study of On-Chain
  Governance.
\newblock In \emph{Financial Cryptography and Data Security}, 2023.

\bibitem[Feichtinger et~al.(2024)Feichtinger, Fritsch, Heimbach, Vonlanthen,
  and Wattenhofer]{feichtinger-et-al-24}
R.~Feichtinger, R.~Fritsch, L.~Heimbach, Y.~Vonlanthen, and R.~Wattenhofer.
\newblock SoK: Attacks on DAOs.
\newblock Technical report, 2024.

\bibitem[Fritsch et~al.(2024)Fritsch, M{\"u}ller, and
  Wattenhofer]{fritsch-et-al-22}
R.~Fritsch, M.~M{\"u}ller, and R.~Wattenhofer.
\newblock Analyzing Voting Power in Decentralized Governance: Who Controls
  DAOs?
\newblock \emph{Blockchain: Research and Applications}, 5\penalty0
  (3):\penalty0 100208, 2024.

\bibitem[Kubinec(2023)]{kubinec-23}
J.~Kubinec.
\newblock Nouns DAO Fork Loses Half Its Treasury in 3 Days.
\newblock \url{https://blockworks.co/news/nouns-dao-treasury-fork-governance},
  2023.

\bibitem[Lloyd et~al.(2024)Lloyd, {{\'O} Broin}, and Harrigan]{lloyd-et-al-24}
T.~Lloyd, D.~{{\'O} Broin}, and M.~Harrigan.
\newblock The On-Chain and Off-Chain Mechanisms of DAO-to-DAO Voting.
\newblock In \emph{The IEEE International Conference on Blockchain}, pages
  649--655. IEEE, 2024.
\newblock \doi{10.1109/Blockchain62396.2024.00095}.

\bibitem[Lloyd et~al.(2025)Lloyd, {{\'O} Broin}, and Harrigan]{lloyd-et-al-25}
T.~Lloyd, D.~{{\'O} Broin}, and M.~Harrigan.
\newblock Mapping Partisan Fault Lines Within DAOs.
\newblock In \emph{The IEEE International Conference on Blockchain}, pages
  38--45. IEEE, 2025.
\newblock \doi{10.1109/Blockchain67634.2025.00015}.

\bibitem[Mickevicius et~al.(2014)Mickevicius, Krilavicius, and
  Morkevicius]{mickevicius-et-al-14}
V.~Mickevicius, T.~Krilavicius, and V.~Morkevicius.
\newblock Analysing Voting Behavior of the Lithuanian Parliament Using Cluster
  Analysis And Multidimensional Scaling: Technical Aspects.
\newblock In \emph{The International Conference on Electrical and Control
  Technologies}, pages 84--89, 2014.

\bibitem[Quan et~al.(2024)Quan, Wu, Deng, and Zhang]{quan-et-al-24}
Y.~Quan, X.~Wu, W.~Deng, and L.~Zhang.
\newblock Decoding Social Sentiment in DAOs: A Comparative Analysis of
  Blockchain Governance Communities.
\newblock In \emph{The IEEE International Conference on Software Quality,
  Reliability, and Security Companion}, pages 216--224, 2024.

\bibitem[Reijers et~al.(2018)Reijers, Wuisman, Mannan, Filippi, Wray, Rae-Looi,
  V{\'e}lez, and Orgad]{reijers-et-al-18}
W.~Reijers, I.~Wuisman, M.~Mannan, P.~D. Filippi, C.~Wray, V.~Rae-Looi, A.~C.
  V{\'e}lez, and L.~Orgad.
\newblock Now the Code Runs Itself: On-Chain and Off-Chain Governance of
  Blockchain Technologies.
\newblock \emph{Topoi: An International Review of Philosophy}, 2018.
\newblock \doi{10.1007/s11245-018-9626-5}.

\bibitem[{Sky (formerly MakerDAO)}(n.d.)]{sky-xx}
{Sky (formerly MakerDAO)}.
\newblock \url{https://sky.money/} and \url{https://archive.today/YVBE3}, n.d.

\bibitem[Sun et~al.(2022)Sun, Chen, Stasinakis, and Sermpinis]{sun-et-al-22-1}
X.~Sun, X.~Chen, C.~Stasinakis, and G.~Sermpinis.
\newblock Voter Coalitions and Democracy in Decentralized Finance: Evidence
  From MakerDAO.
\newblock Technical report, 2022.

\end{thebibliography}

\end{document}